\documentstyle[aps,prl,epsfig,amsmath,amssymb,multicol]{revtex}

\newcommand{\be}{\begin{eqnarray}}
\newcommand{\ee}{\end{eqnarray}}

\begin{document}

\title{A Model for Striped Growth}
\author{Hai Qian and Gene F. Mazenko}
\address{James Franck Institute and Department of
Physics, University of Chicago, Chicago, Illinois 60637}
\date{today}
\maketitle
\begin{abstract}
We introduce a model for describing the defected
growth of striped patterns. This model, while roughly related to the
Swift-Hohenberg model, generates a quite different mixture of defects
during phase ordering. We find two characteristic lengths in the
system: the scaling length $L(t)$, and the average width of the domain
walls. The growth law exponent is larger than the value of $1/2$ found
in typical point defect systems.
\end{abstract}


\begin{multicols}{2}

\section{Introduction}

Our understanding of the growth of striped patterns 
after a temperature quench from an
isotropic state
remains limited \cite{EVG,CM,CB,HSG,BVdc,BVgm,BVgbp,Qian1}.  We do not have the simple
standard scenario of point defect annihilation observed
in systems like the XY model \cite{Qian2} where there is ordering 
ending in a homogeneous phase.  Instead
we have pattern coarsening via a competition between point and line
defects constructed from building blocks of dislocations
and disclinations.
Experiments and numerical simulations for simple models are in agreement
that there is scaling in stripe forming systems governed by a characteristic
length or  growth law, $L(t)\approx t^{x}$, with  an exponent 
$x\approx 1/4$ to $1/3$ \cite{Qian1}.
This exponent is smaller than one would expect from the 
simplest theoretical  treatments which give $x\approx 1/2$.
A convincing theoretical understanding of the values of $x$
for isotropic pattern forming systems is still lacking.  A complication
is that there may not be a well defined value for $x$, but,
as found
in numerical treatments, $x$ may
depend weakly on a control parameter. 

When we turn to the defect structures we find lack of agreement between
models and experiments.
Experimentally the study of stripe formation has taken
a substantial step forward with the work of Harrison \textit{et al}
\cite{Harrison} on
carefully prepared two dimensional diblock copolymer systems which order
into striped systems appearing to be two dimensional smectics.
For quenches into
the appropriate temperature range they find  ordering which proceeds,
in the scaling regime, via the process of the annihilation of a
set of disclination quadrapoles.  They also find that characteristic growth
laws grow in time $L(t)\approx t^{x}$ with exponent $x=1/4$.

The Swift-Hohenburg (SH) model (see Eq. (11) below) \cite{SH} represents a simple
model for producing stripe pattern growth. In a previous paper
\cite{Qian1} we discussed the distribution of defects (grain boundaries
and point defects) which govern the ordering in the SH model in two
dimensions. We found \cite{Qian1} the kinetics dominated by grain
boundaries with a small number of free dislocations and a smaller number
of free disclinations.
Thus the simplest model, the SH model, does not produce 
the defect structure seen in the
experiments. 

In this paper we present an alternative model description,
the nonlinear phase (NLP)  model, for growing stripes. It is
motivated as an approximation to SH model. However the model grows stripes
via a quite different defect structure compared to the SH model, but shares with it  the characteristic feature of grain boundaries.
This model could also, as with the SH model, be constructed by appeal to
symmetry, simplicity and analyticity. 

There appears to be a variety of pathways to equlibrium in these
systems. One may need to appeal to more than just symmetry and
analyticity to obtain a quantitative description of the ordering process
in striped systems.

\section{The Nonlinear Phase Model}

The nonlinear phase model can be obtained as an approximate phase-field
model for the SH model defined by
\be
\partial_t \psi = \epsilon \psi - \left(\nabla^2+1\right)^2 \psi -\psi^3
~~~,
\ee 
where $\psi$ is a real scalar field.
We only consider the zero temperature case in this paper, so there is no noise term
in the above equation. Assume that the solution of the SH model can be written in the
single mode  form \cite{PM79}:
\be
\psi ({\bf x},t)=\psi_{0} \cos (\kappa ( {\bf x},t))
\label{eq:2}
\ee
where the amplitude $\psi_{0}$ saturates quickly at 
the ground state amplitude $\sqrt{4\epsilon/3}$.  Substituting
this ansatz into the SH equation, 
ignoring amplitude fluctuations and higher harmonics, we find
that the coefficient of $ \psi_{0}\sin (\kappa ( {\bf x},t))$
satisfies
\be
-\dot{\kappa}=\nabla_j \left[\nabla^2 Q_j + 2 (1-Q^2)Q_j\right]
\label{eq:3}
\ee
where 
\be
{\bf Q}({\bf x},t)=
\nabla\kappa({\bf x},t)
\label{eq:4}
\ee
is the 
local wavenumber of the stripes. Eq. (3) can  be written as 
\begin{equation}
\dot{\kappa}=-\frac{\delta {\mathcal F}}{\delta \kappa}\ ,
\end{equation}
with
\begin{equation}
{\mathcal F}[\kappa]=\int d^2x\,\left\{\frac{1}{2}(\nabla^2 \kappa)^2+\frac{1}{2}\left[(\nabla
\kappa)^2-1\right]^2\right\}\ .
\end{equation}

If we take the gradient with respect to $\nabla_{i}$,
Eq. (\ref{eq:3}) can
be written in the form
\be
\partial_t Q_i 
=\nabla_{i}\nabla_{j}\frac{\delta {\cal H}_{E}}{\delta Q_{j}}
\label{eq:5}
\ee
where the driving free energy is of the standard Ginzburg-Landau
form
\begin{eqnarray}
{\cal H}_{E}[{\bf Q}]
&=&\int d^{2}x\left[ \frac{1}{2}
\left(\nabla {\bf Q}\right)^{2}+\frac{1}{2}(Q^2-1)^2\right]\nonumber\\
&\equiv& \int d^{2}x~\epsilon ({\bf x})\ .
\label{eq:6}
\end{eqnarray}
Except for the fact the ${\bf Q}$ is longitudinal and
one has an anisotropic diffusion tensor
($\nabla_{i}\nabla_{j}$ rather than $\nabla^{2}\delta_{ij}$),
this would just
be the TDGL model for a conserved vector order parameter.
The model (nonlinear phase model or NLP model) defined by Eqs. (\ref{eq:2}), (\ref{eq:3}) and (\ref{eq:4})
correspond, as we now show, to a well defined stand alone model 
for producing striped patterns.
While we have roughly derived this model from the SH model, its
ordering defect structures are quite different from the SH model.

\section{Numerical Results}

\subsection{Qualitative Results}

Let us begin with a qualitative description of the patterns
generated by this model.  We put Eq. (\ref{eq:3}) on a lattice with
spacing $\Delta r=0.7854$ and time step
$0.01$.  We use the isotropic form for the Laplacian
given by 
\be
\nabla^2 \phi({\bf r},t) \rightarrow
\nabla^2\phi_{ij}=\frac{1}{(\Delta
r)^2}\left[\frac{2}{3}\sum_{NN}+\frac{1}{6}\sum_{NNN}-\frac{10}{3}\right]\phi_{ij}
\ee
where $NN$ and $NNN$ mean the nearest neighbors and next-nearest
neighbors respectively. $\phi$ can be the order parameter $\kappa$ or the
component of ${\bf Q}$. $(i,j)$ is the  lattice-site position.
We run these equations on lattice of various
sizes using initial conditions where
the phase variable $\kappa (i,j)$ is chosen  to have a
value randomly distributed between $-\kappa_{0}$ and 
$\kappa_{0}$ where
$\kappa_{0}=0.01/\sqrt{2}=0.0070711$.

In Fig. 1 we plot those points 
where $\cos\kappa$ is positive for a system, which is on a $512 \times 512$ grid,
after an evolution to time $t= 1000$.
The stripe wavenumber $Q^{2}$ orders very rapidly so that by the time
$t=1000$ we have an interconnected pattern of basically compact objects.
These appear to be nucleated objects with radial symmetry.  

\begin{figure} 
\begin{center}\includegraphics[scale=0.68]{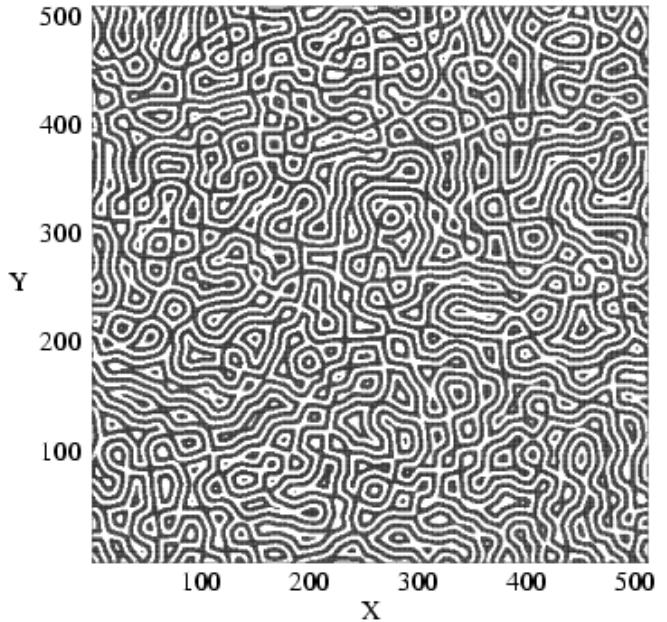}\end{center}
\caption{$\cos\kappa>0$ at $t=1000$ in a $512\times 512$ system.}
\end{figure}

\noindent This is a fairly early time for this system and we see that  we have
generated a set of local donuts.
We have many concentric circles with few direct
paths through the system.  We clearly have layers but they are strongly
bent.

\begin{figure} 
\begin{center}\includegraphics[scale=0.68]{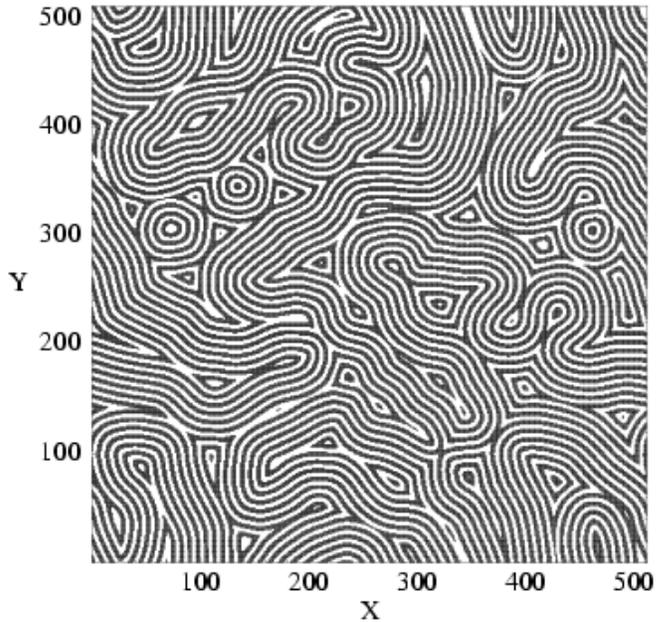}\end{center}
\caption{$\cos\kappa>0$ at $t=5000$ in a $512\times 512$ system.}
\end{figure}

\noindent When we reach a time of 5000 we see that most of the donuts
have opened and stripes are forming and winding through the system.
Clearly we see that the remaining centers of the donuts are serving
as cores for large disclinations.

\begin{figure} 
\begin{center}\includegraphics[scale=0.68]{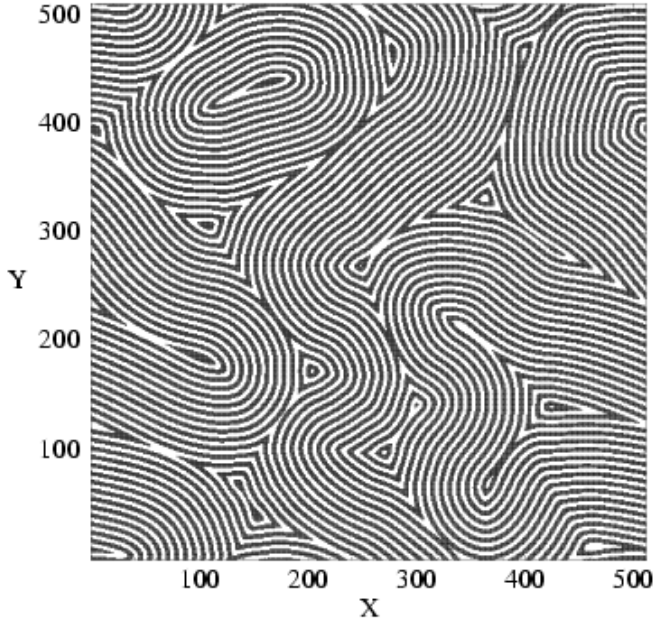}\end{center}
\caption{$\cos\kappa>0$ at $t=15000$ in a $512\times 512$ system.}
\end{figure}

\noindent This pattern coarsens as one moves to $t=15000$, as shown in Fig.3,

\begin{figure} 
\begin{center}\includegraphics[scale=0.67]{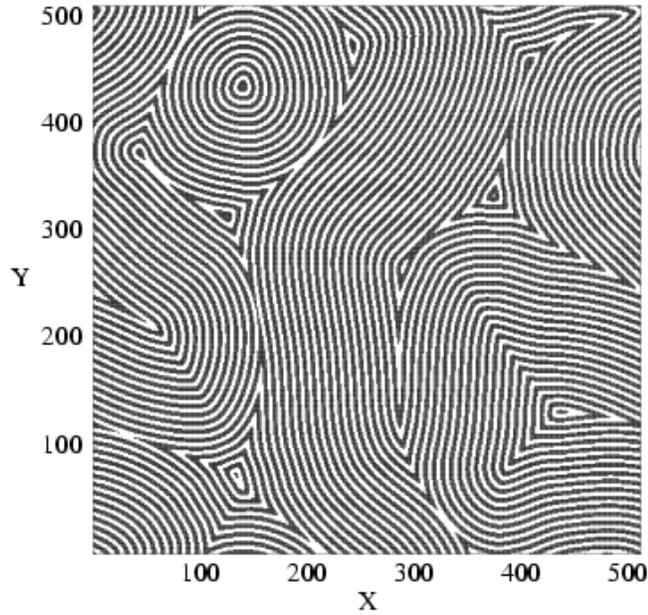}\end{center}
\caption{$\cos\kappa>0$ at $t=25000$ in a $512\times 512$ system.}
\end{figure}

\noindent and Fig. 4 where 
$t=25000$.  In Fig. 4 one sees a target pattern in the upper left.
One expects this to eventually break open.

These patterns can, at later times, be rather completely characterized
by their defect structure.  The defects can be found by 
 looking for positions where the amplitude $Q^{2}$ is significantly
smaller than its ordered value $Q_{0}^{2}=1$.

\begin{figure} 
\begin{center}\includegraphics[scale=0.67]{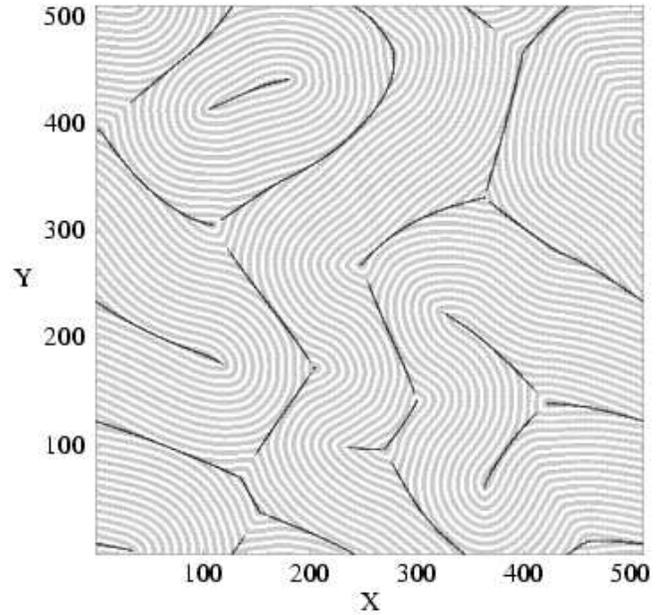}\end{center}
\caption{$\cos\kappa>0$ at $t=15000$ in a $512\times 512$ system. Those
points with a small $Q^2$ are also shown.}
\end{figure}

\noindent In Fig. 5 we plot those points where
$Q^{2}< 0.5$ for the pattern in Fig. 3.  We see that we have a rather complete map of
the defect structure seen in the layer pattern.  The small amplitude lines (SAL) network defines the pattern. 

\begin{figure} 
\begin{center}
\includegraphics[scale=0.6]{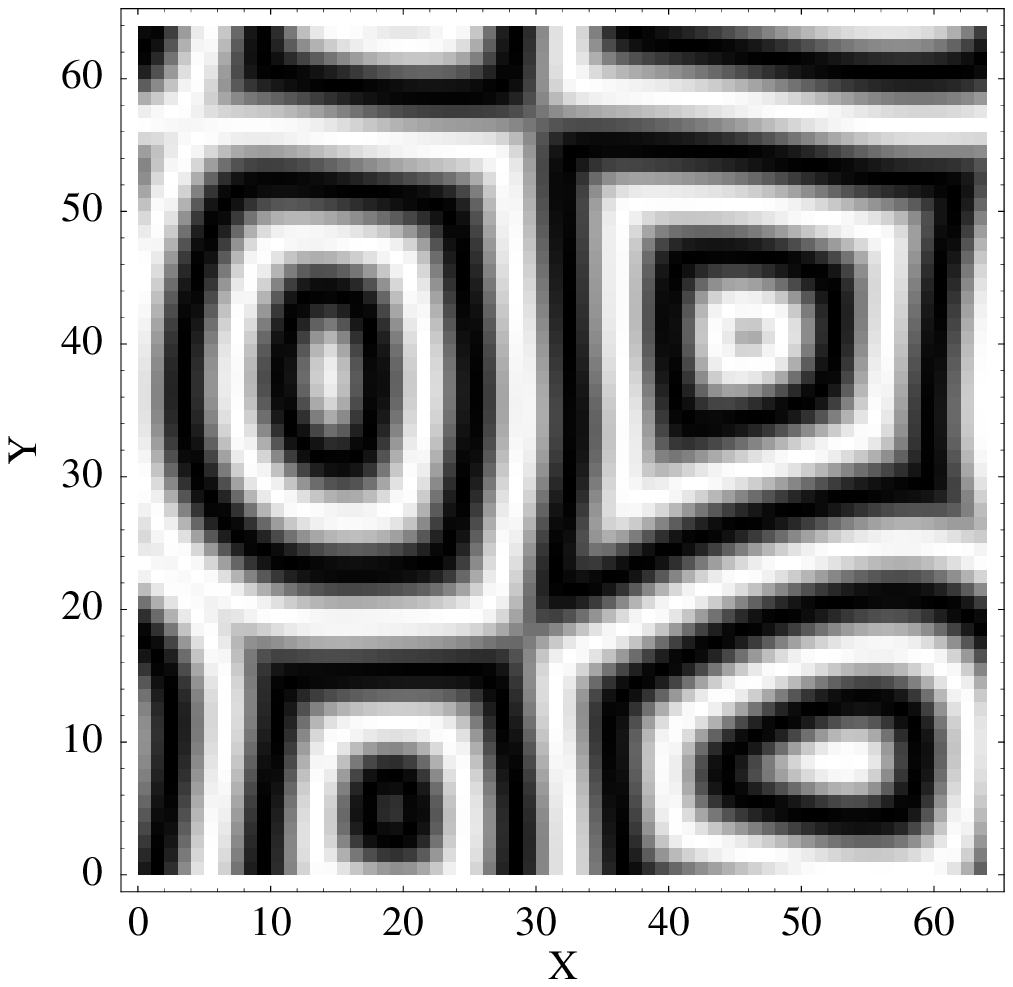}
\includegraphics[scale=0.8]{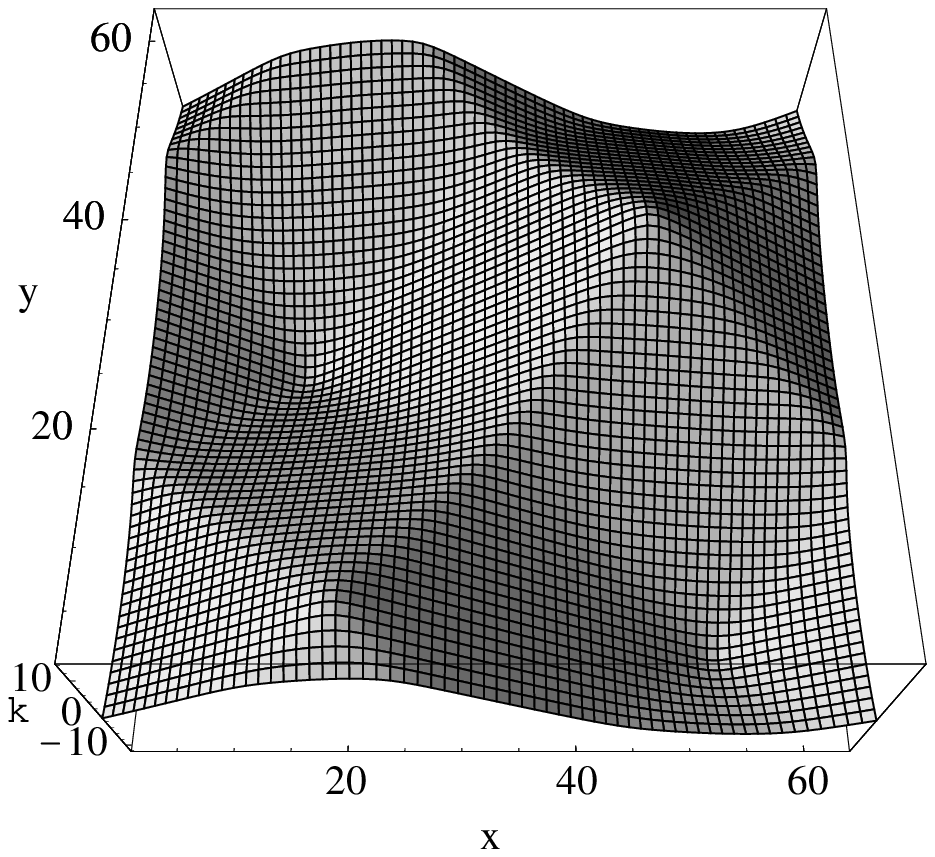}
\end{center}
\caption{The pattern $\cos\kappa$ and $\kappa({\bf x})$ for a $64\times 64$ system
at $t=800$. }
\end{figure}

\noindent In Fig. 6 we plot the pattern $\cos\kappa>0$ and $\kappa({\bf x})$ for a $64\times 64$ system
at $t=800$. Clearly the peaks and anti-peaks correspond to the target
centers of the pattern $\cos\kappa$, and the edges correspond to the SAL
network. We also observe that the peaks and anti-peaks correspond to the $+1$
vortices in the ${\bf Q}$ field, and on some of the edges there is a $-1$ vortex of
the ${\bf Q}$ field, as is shown in Fig. 7. The numbers of $+1$ and $-1$
vortices are equal, while the numbers of the peaks (anti-peaks) and the
edges are usually not equal, so not every edge has a $-1$ vortex on it.

\begin{figure} 
\begin{center}\includegraphics[scale=0.85]{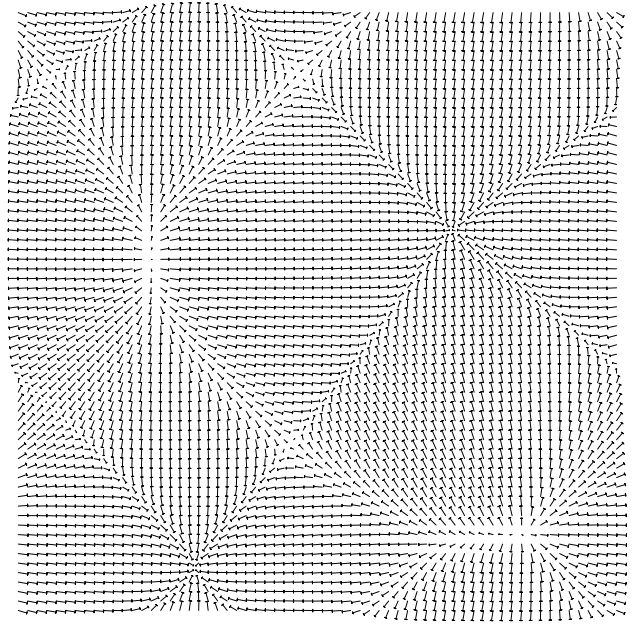}\end{center}
\caption{The ${\bf Q}$ field for the pattern in Fig. 6. The vector on
each site is shown.}
\end{figure}

The $+1$ vortices are compact but the $-1$ vortices are not. This can be
seen in Fig. 7, where some of the  $-1$ vortices occupy entire edges. So the
defects in the system are more like a domain-wall network rather than a
set of vortices. The domain walls correspond to the edges in Fig. 6.

\subsection{Quantitative Results}

At long times after a quench, 
phase ordering systems typically enter a scaling regime with a single
characteristic length $L$. Here we check scaling for the correlation function $C_{\kappa}({\bf r},t)=\langle
\kappa({\bf x}+{\bf r},t)\,\kappa({\bf x})\rangle$. In Fig. 8, we show
this correlation function at 8 different times. The data is scaled so
that we can see whether it obeys a scaling law. The correlation
length can be extracted from $C_{\kappa}(r_0,t)/C_{\kappa}(0,t)=1/2$ where
$r_0\propto L$. The result is shown in Fig. 9. Since $|\nabla \kappa|\sim
$constant, we expect that $\kappa\sim L$ and $\displaystyle
C_{\kappa}(0,t)=\langle\kappa^2\rangle\sim L^2$. Thus we expect the
scaling form 
$C_\kappa(r,t)=L^2\,F(r/L)$. The direct measurement of
$\langle\kappa^2\rangle$ is shown in Fig. 10. 
From Fig. 9, we get $L\sim t^{0.60}$. From Fig. 10, we also obtain $L\sim
t^{0.60}$.

\begin{figure} 
\begin{center}\includegraphics[scale=0.31]{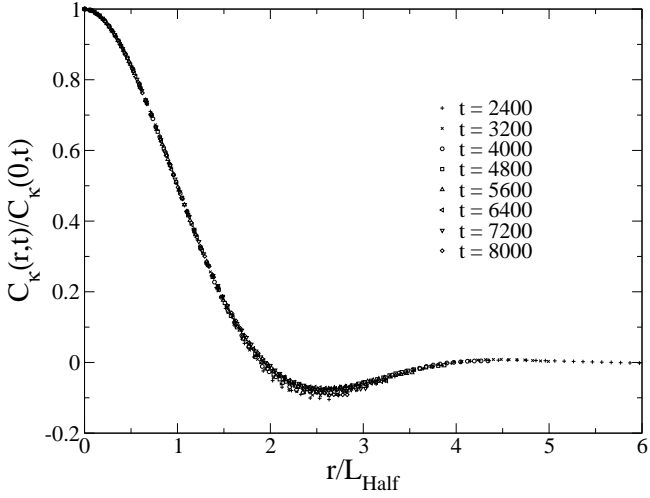}\end{center}
\caption{The correlation function $C_{\kappa}(r,t)$ for the $256\times 256$
system is plotted, for 8
different times, in scaling form. The distance $r$ is scaled
by $L_{Half}=r_0$ given in Fig. 9. At least at small $r$, the
function has a scaling form. The data is averaged over 41 runs.}
\end{figure}

\begin{figure} 
\begin{center}\includegraphics[scale=0.31]{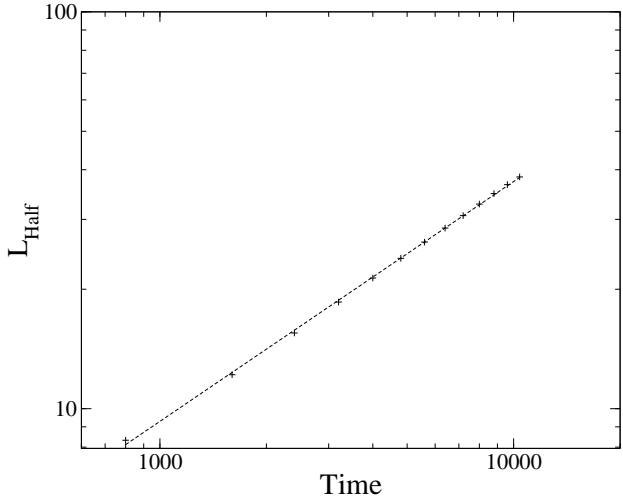}\end{center}
\caption{We extract the correlation length from
$C_{\kappa}(r_0,t)/C_{\kappa}(0,t)=1/2$. $r_0=L_{Half}$ is proportional to $L$. We
find $L\sim t^{0.60}$. Averaged over 41 runs.}
\end{figure}

\begin{figure} 
\begin{center}\includegraphics[scale=0.31]{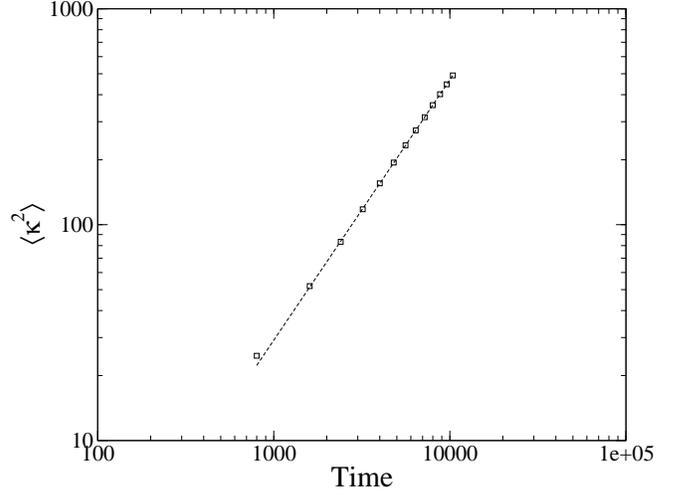}\end{center}
\caption{$\langle \kappa^2\rangle \propto t^{1.20}$. So $L^2 \sim t^{1.20}$ and
$L\sim t^{0.60}$. Averaged over 41 runs.}
\end{figure}

\noindent We measure next the
number of defects in the system as a function of time after the quench.
In Fig. 11  we plot the density of sites where $Q^{2}< 0.4$ in the
scaling regime.  This curve
is well fit by
\be
n_{v}=a\,t^{-n}
\ee
with $a=3.89$, $n=0.506$. Since the total number of the
defects is proportional to the area of domain walls, the number
density is proportional to $w\cdot L/L^2=w\cdot L^{-1}$, where $w$ is the
average width of the domain walls. Thus we estimate $L/w\sim 0.5$. 

\begin{figure} 
\begin{center}\includegraphics[scale=0.31]{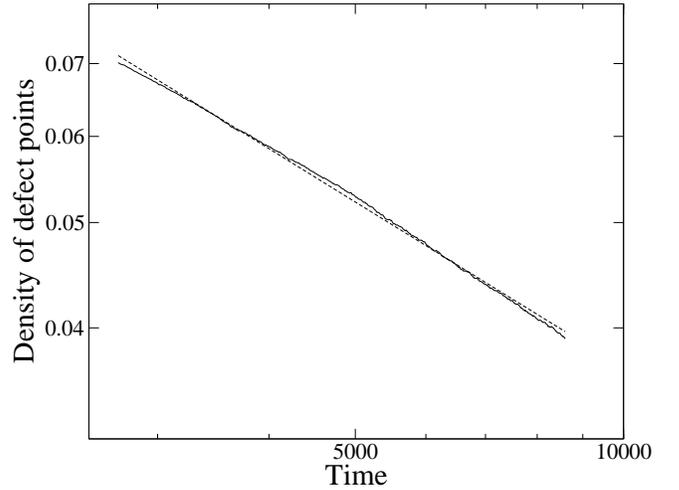}\end{center}
\caption{Density of the defect points where $Q^{2}< 0.4$. Averaged over
38 runs.}
\end{figure}

\noindent It will turn out to be necessary to allow the width $w$ to be
a function of time.

We need an independent method for determining $L$ or $w$. We do this by
employing another method to measure the
length of the defect network directly. Basically this is a
coarse-graining method. We put the defect network on a lattice with
lattice spacing $8\Delta r$ and count the number of sites it
occupies. This number reflects the length of the network much better
than the previous measurement. Other sizes of the lattice spacing can
also be used as long as the grid size is larger than the width of the
domain walls and much smaller than their lengths. The length density
of the network is
shown below and can be fit to $\sim t^{-0.60}$. Since the length density is proportional to
$L/L^2=L^{-1}$, we get $L\sim t^{0.60}$, which is consistent with the
measurement of the correlation function.

\begin{figure} 
\begin{center}\includegraphics[scale=0.31]{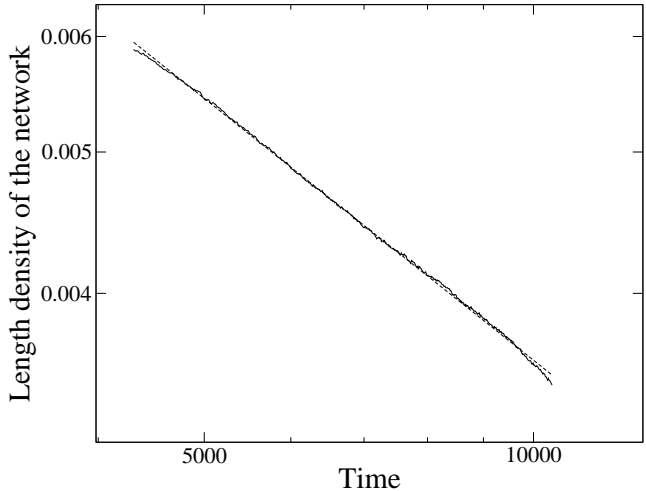}\end{center}
\caption{Length density of the defect points where the length of the
network is in fact the number of sites that the network occupies in the
lattice with larger grid. The dashed line is proportional to $t^{-0.60}$. Averaged over
38 runs.}
\end{figure}

We can then conclude that the average width of the domain walls $w\sim
t^{0.10}$. In the time regime we study, the width $w$ is about $2\sim
4$. We don't think $w$ will increase for ever, it may stop increasing at
some later stage. But before that can happen in our system, finite size
effect enters.

The growth laws for $L$ and $w$ can explain the growth exponents of other
quantities. We give two examples below: $\langle Q^{2}\rangle$ and the energy
density $\epsilon$ in Eq. (8).

The ordering
of the field ${\bf Q}$ is characterized by the average over all
sites of ${\bf Q}^{2}(i,j)$.  We obtain the results shown in Fig. 13.
It can be seen in the figure that there are two regimes where
the data can be fit to a form
\be
\langle Q^{2}\rangle=\frac{1}{a+bt^{-n}}
~~~.
\ee

\begin{figure} 
\begin{center}\includegraphics[scale=0.31]{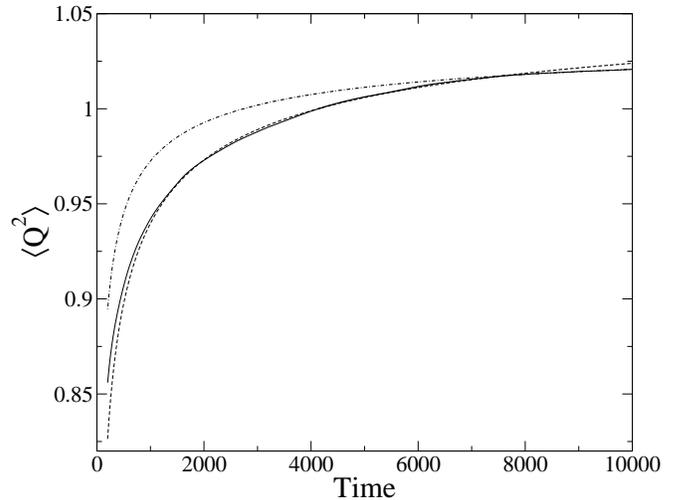}\end{center}
\caption{$\langle Q^2\rangle$ v.s. $t$. Averaged over 20 runs.}
\end{figure}

\noindent In the shorter time regime $1000\leq t\leq 7000$ the data is fit with
with $a=0.930$, $b=3.15$, and $n=0.45$.  In the longer time regime
$7500\leq t \leq 10000$ we have the fit
$a=0.950$, $b=2.40$, and $n=0.511$.  In both fits the exponent
$n$ is near $1/2$.  We also see that in both fits the final value
of $Q^{2}$ is larger than 1.  Clearly more work needs to be done to
establish the nature of this cross over.  It seems likely that the
final asymptotic value of $Q^{2}$ is greater than one due to 
finite size effects and the freezing of grain boundaries for long times.

In this system, $Q^2$ is approximately $1$ away from the domain-wall
network and small on the domain walls with an average $c<1$. Because the average
area $s$ of the domain walls in one domain is proportional to 
$wL$, and the domain's area is $A\sim L^2$, we have 
\begin{equation}
\langle Q^2 \rangle = \frac{1}{A}\left[(A-s)\cdot 1+s\cdot
c\right]=1-(1-c)\cdot \frac{s}{A}\ ,
\end{equation}
where $s/A \propto wL^{-1}$. So the average of $Q^2$ has the form of
$\langle Q^2\rangle-Q_0^2\sim w/L$. Thus we can identify $L/w\sim t^{0.5}$.

\begin{figure} 
\begin{center}\includegraphics[scale=0.31]{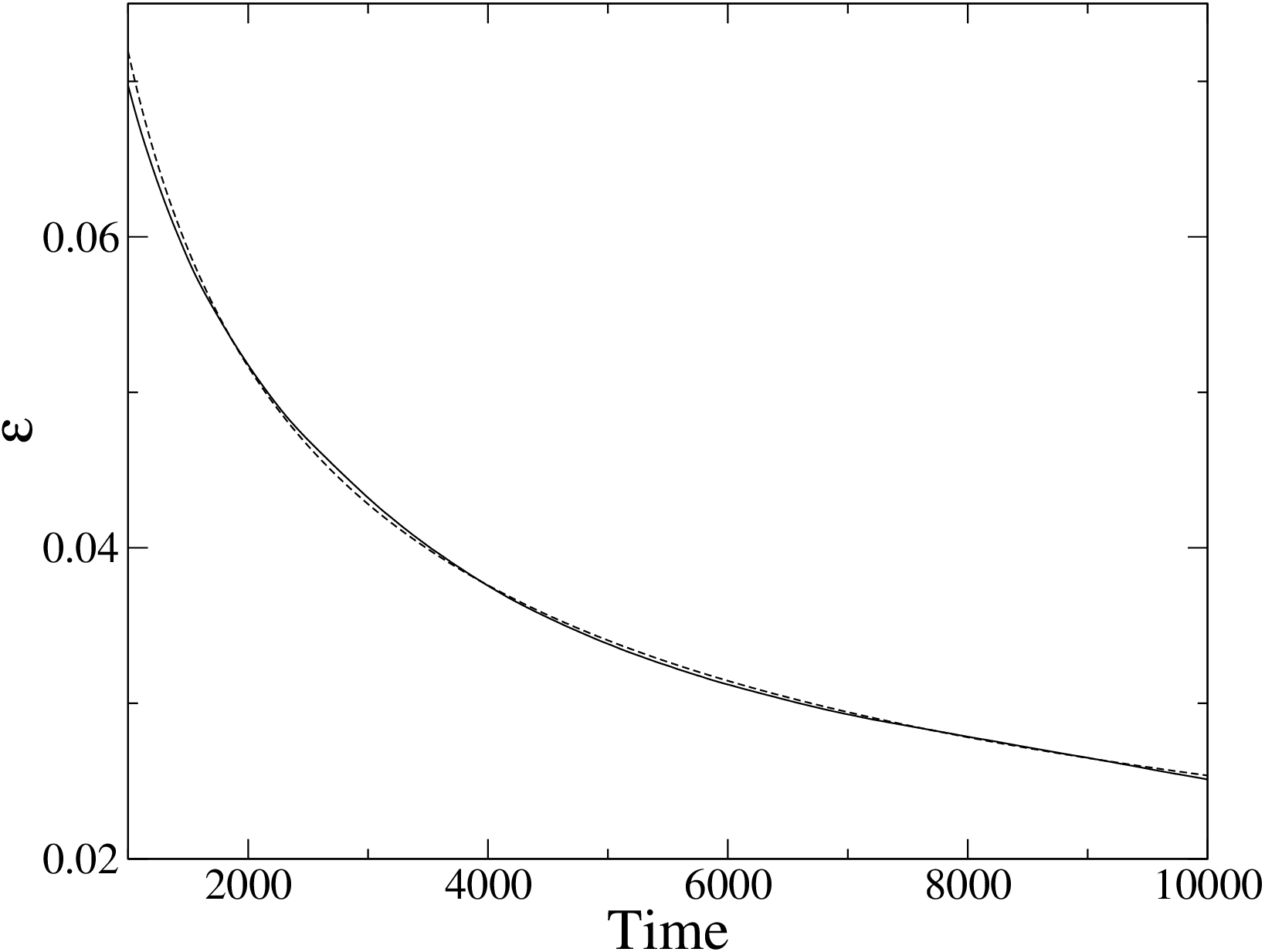}\end{center}
\caption{The energy density of the system. Averaged over 20 runs.}
\end{figure}

In Fig. 14 we show the path toward equilibration of the effective
energy density $\epsilon $ defined by Eq.(\ref{eq:6}).  This
result seems to be in agreement with that for $Q^{2}$.  We have a good
fit to
\be
\epsilon =a+bt^{-n}
\ee
with $a=0.0056$, $b=2.526$ and $n=0.5288$. Since the energy above the
ground state should be proportional to the area of the domain walls,
the energy density is proportional to $wL/L^2=wL^{-1}$. Again we get
$L/w\sim t^{0.5}$.

\section{Discussion}

In the NLP model we know that there are analytic vortex solutions related
to those for the XY model.
This would seem to favor coarsening via a set of isolated vortices 
which pair up
and annihilate.  Instead we find large $-1$ vortices forming a domain
wall network. 

This system does not appear to generate  dislocations and, as such, is quite
different from the SH model which has a significant density of dislocations.
The nonlinear phase model, on a larger scale, is growing targets and disclinations.

The NLP model introduced here helps to deepen our feeling that we do not
have a good understanding of the general mechanism of stripe
formation. In the SH model and the appropriate experiments the ordering
is slow compared with the simple model of point defect annihilation. In
this NLP model the ordering is ``faster'' than the simple model. 

In this model we find two characteristic lengths, $L$ and $w$, which
together explain the
different exponents we observed. In the SH model, we also observed
different exponents \cite{Qian1}. Our guess is that in SH model there are also more
than one characteristic length.

In conclusion we see that models with the same symmetries can vary
significantly in the defect structures produced during ordering. The
search continues for models where one can dial the relative abundance  of grain
boundaries and free dislocations and disclinations. The ultimate goal is
to match the models with a given experimental system.

\vspace{5mm}

Acknowledgments: This work was supported by the National Science Foundation
under Contract No. DMR-0099324.

\end{multicols}

\end{document}